\documentclass[11pt,showpacs,pop,aip,preprint]{revtex4-1}

\usepackage{graphicx}
\usepackage[usenames]{color} 
\usepackage[normalem]{ulem}
\usepackage{hyperref}
\definecolor{solarRed}{RGB}{220,50,47} 
\definecolor{orange}{RGB}{220,140,10} 
\definecolor{purple}{RGB}{100,0,160} 
\definecolor{dkred}{RGB}{160,60,50} 

\newcommand{\brac}[1]{\ensuremath{\left(#1\right)}}
\newcommand{\eqn}[1]{eq.~\ref{#1}}

\newcommand{\fig}[1]{figure~\ref{#1}}
\newcommand{\figs}[1]{figures~\ref{#1}}
\newcommand{\Fig}[1]{Figure~\ref{#1}}
\newcommand{\Ref}[1]{{\setcitestyle{numbers}Ref.~\cite{#1}\setcitestyle{super}}}
\newcommand{\order}[1]{\ensuremath{\mathcal{O}\brac{#1}}}

\newcommand{\phixt}{\ensuremath{\phi\brac{x,\theta}}}
\newcommand{\thz}{\ensuremath{{\theta_0}}}
\newcommand{\game}{\ensuremath{\gamma_E}}
\newcommand{\rhoi}{\ensuremath{\rho_i}}

\newcommand{\nqp}{\ensuremath{nq^\prime}}
\newcommand{\etai}{\ensuremath{\eta_i}}
\newcommand{\shat}{\ensuremath{\hat{s}}}
\newcommand{\kth}{\ensuremath{k_\theta}}

\newcommand{\xitt}{\ensuremath{\xi\brac{\theta,\thz}}}
\newcommand{\atz}{\ensuremath{A\brac{\thz}}}
\newcommand{\iso}{\textit{isolated}}
\newcommand{\gen}{\textit{general}}
\newcommand{\umx}{\ensuremath{u_m\brac{x}}}

\begin{document}

\title{
Structure of Micro-instabilities in Tokamak Plasmas: Stiff Transport or Plasma Eruptions?
}


\author{D~Dickinson}
\email{dd502@york.ac.uk}
\affiliation{
	York Plasma Institute, 
	Department of Physics, 
	University of York, 
	Heslington, 
	York, 
	YO10 5DD, 
	UK
	}
\affiliation{
	EURATOM/CCFE Fusion Association, 
	Culham Science Centre, 
	Abingdon, 
	Oxon, 
	OX14 3DB, 
	UK
}
\author{C~M~Roach}
\affiliation{
	EURATOM/CCFE Fusion Association, 
	Culham Science Centre, 
	Abingdon, 
	Oxon, 
	OX14 3DB, 
	UK
}
\author{J~M~Skipp}
\author{H~R~Wilson}
\affiliation{
	York Plasma Institute, 
	Department of Physics, 
	University of York, 
	Heslington, 
	York, 
	YO10 5DD, 
	UK
	}



\begin{abstract}
Solutions to a model 2D eigenmode equation describing micro-instabilities in tokamak plasmas are presented that demonstrate a sensitivity of the mode structure and stability to plasma profiles.
In narrow regions of parameter space, with special plasma profiles, a maximally unstable mode is found that balloons on the outboard side of the tokamak.
This corresponds to the conventional picture of a ballooning mode.
However, for most profiles this mode cannot exist and instead a more stable mode is found that balloons closer to the top or bottom of the plasma.
Good quantitative agreement with a 1D ballooning analysis is found provided the constraints associated with higher order profile effects, often neglected, are taken into account.
A sudden transition from this general mode to the more unstable ballooning mode can occur for a critical flow shear, providing a candidate model for why some experiments observe small plasma eruptions (Edge Localised Modes, or ELMs) in place of large Type I ELMs.
\end{abstract}

\maketitle

Micro-instabilities in magnetised plasmas are those with a characteristic length scale across magnetic field lines comparable to the ion Larmor radius, $\rhoi$.
The particle drifts play a key role in the dynamics, and therefore often characterise their growth rate, which is weak compared to Alfv\'{e}nic modes.
While these micro-instabilities are relatively benign they are nevertheless important because they drive turbulence that degrades plasma confinement.
Understanding turbulence and its influence on confinement is one of the key challenges facing magnetically confined fusion plasmas. 

In this Letter, we consider the generic linear properties of micro-instabilities that drive turbulence in a toroidal magnetic confinement device, such as a tokamak.
Our new 2D calculations of their eigenmode structure are compared with analytic `ballooning formalism' approaches that reduce the problem to 1D \cite{taywilcon,dewarPPCF}.
Ballooning models have identified two types of mode structure: (1) an \iso{} mode is the most unstable and is related to the conventional ballooning mode of ideal magnetohydrodynamics (MHD) \cite{cht, glasser, dewar, lee}, but only exists in very special circumstances, and (2) a more stable \gen{} mode, that exists throughout the plasma.
Both of these mode structures are found in our 2D eigenmode calculations, providing first quantitative numerical tests that confirm ballooning theory.
Our results suggest a possible mechanism for a sudden transition from benign micro-instabilities driving turbulent transport to stronger instabilities that could release small filamentary eruptions and locally collapse the profiles.
This mechanism may form the basis of a model for small Edge Localised Modes (ELMs) in tokamaks \cite{oyama}.

While our results are generic to any micro-instability in a tokamak plasma, it is helpful to illustrate the analysis with a particular model.
Thus, we consider a large aspect ratio, circular cross section tokamak equilibrium, with flux surfaces labelled by the minor radius, $r$.
We consider only electrostatic fluctuations, and assume an adiabatic electron response.
We solve the gyrokinetic equation for the ion distribution function, which is expanded assuming that the ion transit and drift frequencies are small compared to the mode frequency.
For a wavelength across magnetic field lines that is larger than the ion Larmor radius, quasi-neutrality provides the following equation for the perturbed electrostatic potential, $\phi$ \cite{ct}:
\begin{widetext}
	\begin{equation}
		\left[
			\rhoi^2 \frac{\partial^2 }{\partial x^2}
		 - 	\frac{\sigma^2}{\omega^2}
		 	\brac{
		  		\frac{\partial}{\partial \theta}
		  	 + 	i \kth \shat x
		  	 	}^2
		 - 	\frac{2 \epsilon_n}{\omega}
		 	\brac{
		 	 	\cos \theta 
		 	 +	\frac{i \sin \theta}{\kth }
		 	 	\frac{\partial}{\partial x}
		 	 	}
		 -	\frac{\omega -1}{\omega + \etai} 
		 -	\kth^2 \rhoi^2
		 \right]
		 \phixt = 0
	\label{model-eqn}
	\end{equation}
\end{widetext}
where $\sigma =\epsilon_n/(q \kth \rhoi)$, $\shat$ is the magnetic shear, $q$ is the safety factor, $\epsilon_n = L_n/R$, $L_n$ is the density gradient scale length, $R$ is the major radius, $\omega$ is the complex mode frequency normalised to the electron diamagnetic frequency, $\kth$ is the poloidal wave number, $\etai$ is the ratio of density to ion temperature gradient length scales and we have assumed equal electron and ion temperatures.
The poloidal angle, $\theta$ is defined so that $\theta=0$ at the outboard mid-plane and $x=r-r_s$ is the distance from the rational surface at $r=r_s$.
The model is valid in the core relevant limit $\etai \gg 1$ which recovers the ion temperature gradient mode (ITG), which is our focus here.

The 2D eigenmode equation can be conveniently solved by adopting a Fourier transform representation of $\phi$ \cite{zhang}:
\begin{equation}
\phixt = 
	\!\int_{\!\textrm{\,-}\infty}^{\infty}{
		\! \mkern-18mu \atz \xitt
		\exp\left[
			i\nqp x (\thz-\theta)
			\right] 
			\mathrm{d}\,\thz
		}
\label{ft}
\end{equation}
so that $\nqp \thz $ can be interpreted as a radial wave-number at $\theta=0$ ($n$ is the toroidal mode number, $q^\prime = dq/dr$ and $\nqp=\kth\shat$).
The amplitude factor $\atz$ is assumed to vary faster with $\thz$ than $\xitt$.
Substituting \eqn{ft} into \eqn{model-eqn} and following a procedure which extends that set out in \Ref{taywilcon} (and analysed in detail in \Ref{dewarPPCF}) we can then reduce \eqn{model-eqn} to a sequence of 1D ordinary differential equations by expanding in $n$, which is assumed to be large.
This yields to leading order the ballooning equation for $\xitt$:
\begin{widetext}
	\begin{equation}
		\left[
			\frac{\sigma^2}{\omega^2}
			\frac{d^2}{d \theta^2}
		+   \kth^2 \rhoi^2 \shat^2\brac{\theta-\thz}^2
		+	\frac{2\epsilon_n}{\omega} \left[
					\cos \theta
				+	\shat\brac{\theta-\thz}\sin\theta
				\right]
		+	\frac{\omega-1}{\omega+\etai} -\kth^2 \rhoi^2
		\right] \xitt
		=0
	\label{1D_Model}
	\end{equation}
\end{widetext}
providing an eigenvalue condition that relates $\omega$, $\theta_0$ and $x$, written as $\omega=\Omega \brac{x,\thz}$.
$\Omega\brac{x,\thz}$ is obtained by solving \eqn{1D_Model} numerically over all $\thz$ for the range of interest in $x$, noting that the equilibrium parameters vary slowly with $x$.
Restricting consideration to small $x$ (anticipating modes localised about $r=r_s$), we can Taylor expand $\Omega = \Omega_0 \brac{\thz} + \Omega_x\brac{\thz} x +[\Omega_{xx} (\thz)/2]x^2+\cdots$.
Multiplying by $\phi$ and transforming into Fourier space, it is straightforward to show that $x \phi$ is proportional to $(i/\nqp) dA/d\thz$ and $x^2 \phi$ to $-1/(\nqp)^2 d^2A/d\thz^2$, so we have:
\begin{equation}
\frac{\Omega_{xx}\brac{\thz}}{2 \brac{\nqp}^2}
	\frac{d^2 A}{d \thz^2}
-\frac{i \Omega_x\brac{\thz}}{\nqp}\frac{dA}{d \thz}
+\left[
	\omega - \Omega_0\brac{\thz}
\right] A 
=0
\label{env}
\end{equation}
It follows from \eqn{1D_Model} that $\xitt= \xi\brac{\theta +2 l \pi, \thz + 2 l \pi}$ for any integer $l$, so from \eqn{ft} $\phi$ is periodic in $\theta$ provided $A$ is periodic in $\thz$.
This provides the boundary condition that determines $\omega$ as an eigenvalue of \eqn{env}.
Solving for $\atz$, together with $\xitt$, provides the 2D eigenfunction $\phixt$.

Analytically, we can deduce two types of solution from \eqn{env}, which are the `\iso{}' and `\gen{}' modes of \Ref{taywilcon}.
The \iso{} mode exists at the special radial location where $\Omega_x = 0$; this is the classic ballooning mode originally derived for ideal MHD \cite{cht, glasser, dewar, lee}.
$A\brac{\thz}$ is highly localised around the $\thz$ value where $\Omega_0\brac{\thz}$ is stationary, say $\thz=\theta_m$.
Expanding about that position, \eqn{env} becomes a Hermite equation, with solution:
\begin{equation}
\atz = 
	\exp \left[ 
		-\frac{\nqp}{2} \brac{
			\frac{\Omega_{\thz\thz}}
			{\Omega_{xx}}
		}^{1/2}
		\brac{\thz-\theta_m}^2
	\right]
\end{equation}
where the sign of the square root is chosen to give a bounded solution in $\thz$.
$\atz$ has a width $\sim n^{-1/2}$, justifying our Taylor expansion of $\Omega_0 (\theta_0)$ about $\theta_0=\theta_m$.
The eigenvalue condition provides:
\begin{equation}
\omega = 
	\Omega_0\brac{\thz=\theta_m} 
	-\frac{\brac{\Omega_{xx} \Omega_{\thz\thz}}^{1/2}}{2\nqp} +\order{n^{-2}}
\end{equation}
Note that the \iso{} mode growth rate is determined from the 1-D ballooning equation by evaluating the eigenvalue at the $x$ and $\thz$ values which maximise the growth rate.
Furthermore, if $\atz$ is highly localised, the Fourier transform of \eqn{ft} will be dominated by the region around $\thz = \theta_m$.
This, together with the fact that $\xitt$ peaks close to $\theta= \thz$, leads to an expression for the potential, $\phi$ which peaks at $\theta =\theta_m$.
For our up-down symmetric model $\theta_m=0$, and an \iso{} mode will be localised on the outboard side of the tokamak. 

To summarise, the \iso{} mode exists when $\Omega_x=0$; one selects the value of $\thz$ to maximise the growth rate, and the mode is localised about $\theta=\thz$, which is often at the outboard mid-plane.
While this is intuitive from a physics point of view, if $\Omega_x \neq 0$ (which is usually the case), the constraints of the higher order theory do not allow such a mode to exist, as we now discuss.

In general, the first order derivative of \eqn{env} cannot be neglected.
Often $\Omega_x$ is complex, and it can only be neglected when its real \textit{and} imaginary parts vanish at the same value of $x$; hence the \iso{} mode only exists under very special situations.
In the more general case with finite $\Omega_x$ we can neglect the $\Omega_{xx}$ term of \eqn{env}.
Dividing the remaining terms by $A \Omega_x$ and integrating over a full period in $\thz$, we derive the eigenvalue condition $\omega = \left\langle \Omega_0 \Omega_x^{-1} \right\rangle/\left\langle \Omega_x^{-1} \right\rangle$, where angled brackets denote averaging over $\thz$.
Substituting this value of $\omega$ into \eqn{env} and integrating yields the required periodic expression for $\atz$.
One finds that $\atz$ is still localised in $\thz$ provided $\Omega \brac{x , \thz}$ is complex, but now around the value of $\thz$ where $\omega=\Omega_0\brac{\thz}$; in our simple model this position is typically at the top or bottom of the tokamak\footnotemark{}, corresponding to $\thz=\pm\pi/2$.
This \gen{} mode is more stable than the \iso{} mode, with a growth rate which is the average of $\Omega_0$ over $\thz$ rather than the maximum, and the mode sits away from the outboard mid-plane.
Due to the difference in stability, the \gen{} mode will have a higher critical gradient than the \iso{} mode.

To provide quantitative numerical tests of this theory, we solve \eqn{1D_Model} for a specific parameter set: $n=50$, $\shat=2$, $\kth \rhoi =0.33$, $R/a=10$, $q\brac{r=r_s}=1.8$ and $q\brac{r}=3.45\brac{r/a}^2$.
We further assume that all parameters in \eqn{model-eqn} are independent of $x$ except for $\etai$.
In this system an \iso{} mode should exist if $\etai$ has a maximum in $x$  (as we shall illustrate), while a linear $\etai$ profile should yield a \gen{} mode.
For the various $\etai$ profiles chosen we fix $\etai=5$ at $r=r_s$ giving the same local eigenvalue at $x=0$, $\Omega\brac{x=0,\thz}$, for each case.
While $q$ is held constant in the co-efficients of \eqn{model-eqn}, $\shat\neq0$ is required to provide a distribution of rational surfaces across the minor radius.

%
\begin{figure}[htb]
    \centering
    \resizebox{1.0\columnwidth}{!}{
		\includegraphics[width=0.5\columnwidth]
		 	{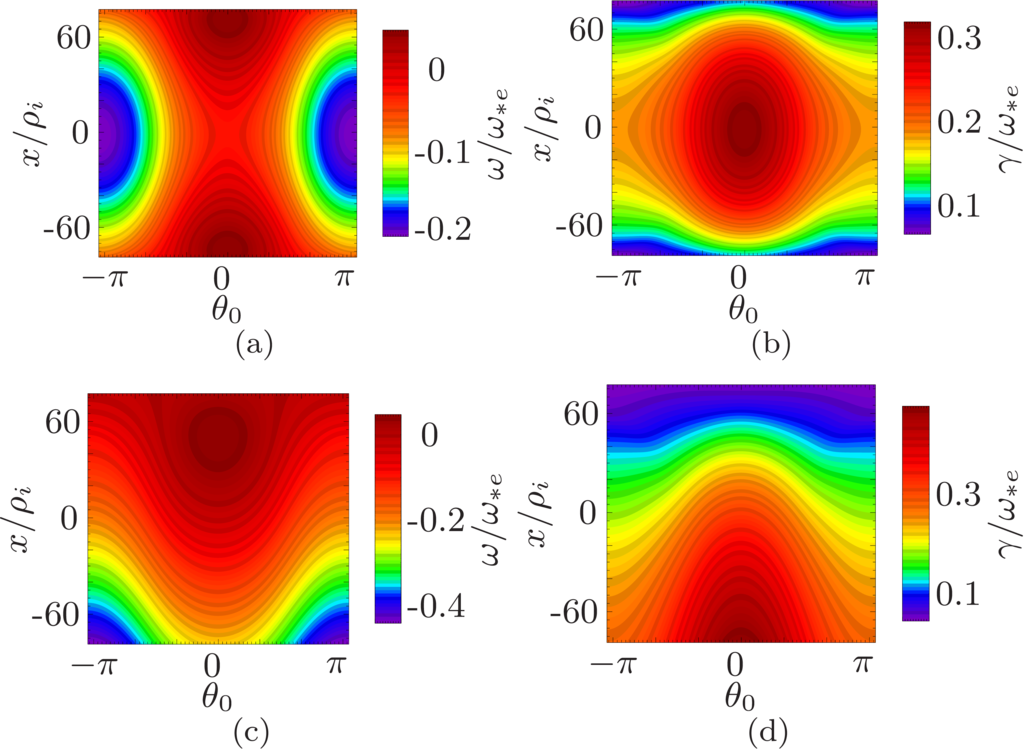}
    	}
 	\caption[]{
 	Contours of the local frequency (a,c) and growth rate (b,d) as a function of $x$ and $\thz$ derived from the 1D ballooning code solution of \eqn{1D_Model} for the peaked and linear $\etai$ profiles respectively.
 	}
	\label{contour}
\end{figure}
%

\Fig{contour} shows $\Omega\brac{x,\thz}$ from solving \eqn{1D_Model} for two types of $\etai$ profile: a quadratic profile peaking near the rational surface \fig{contour}(a, b), and one which decreases linearly in $x$ \fig{contour}(c, d).
Both exhibit $\Omega_0 \sim \cos \thz$, while $\Omega_x$ and $\Omega_{xx}$ are approximately independent of $\thz$.

For the peaked $\etai$ profile, $\Omega_x=0$ at $r=r_s$ and we expect an \iso{} mode.
(This is a special situation: if we were to include radial variations in the other equilibrium parameters of our model then, even for the peaked $\etai$ profile, the real and imaginary parts of $\Omega$ would not be stationary at the same $x$, and no position would exist where $\Omega_x=0$.)
The complex mode frequency from the 1D ballooning procedure is $\omega = -0.025 + i 0.319$. 

With the linear $\etai$ profile $\Omega\brac{x,\thz}$ no longer has a maximum at $x=0$.
Employing the averaging procedure relevant to the \gen{} mode, we obtain $\omega =-0.108 +i0.230$, revealing a substantially more stable situation than for the peaked $\etai$ profile.
Recall that in both cases the local equilibrium parameters are identical at $x=0$.
Solving \eqn{env} for $\atz$ with this \gen{} mode eigenvalue, we find that $\ln \atz \sim \sin \thz$ (with a complex coefficient), so $\atz$ is strongly peaked around $\thz = \pi /2$.
For our circular geometry, this results in a mode structure for $\phixt$ that is localised at the top of the plasma. 
Whether the localisation is at the top or bottom of the plasma depends upon the sign of $\Omega_\thz/\Omega_x$.

\begin{figure}[htb]
    \centering
    \resizebox{1.0\columnwidth}{!}{
			\includegraphics[width=0.45\columnwidth]
		 	{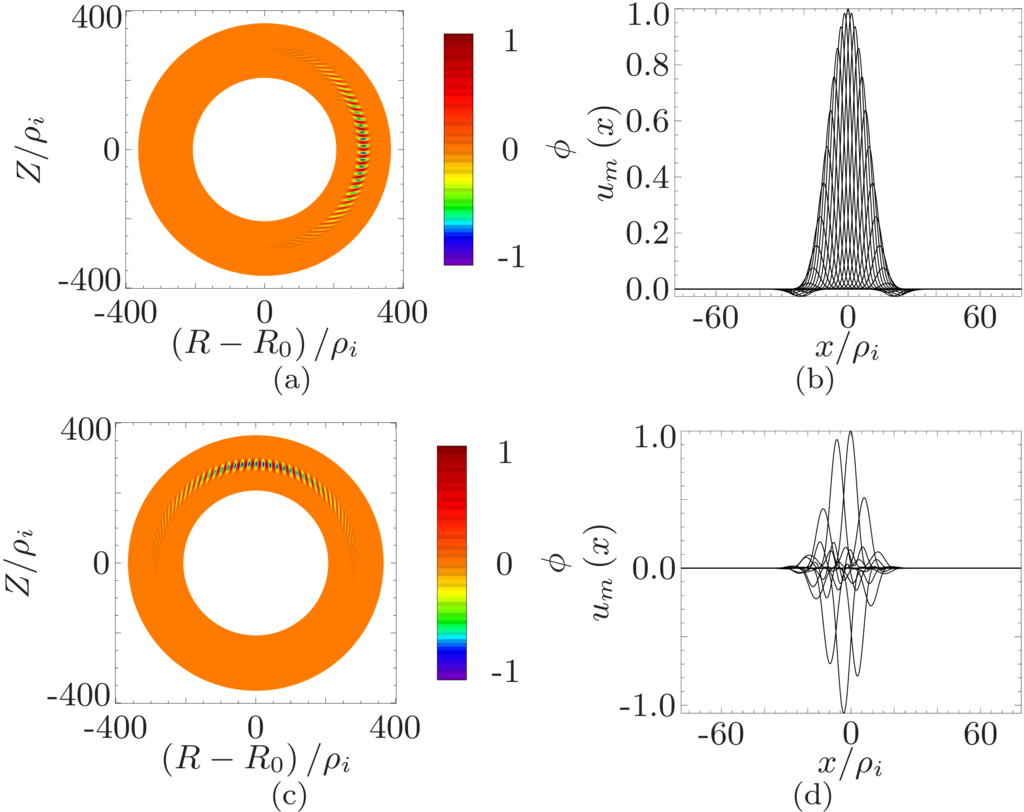}
    	}
 	\caption[]{
 2D eigenmode structure for the potential $\phi$ (a,c) in the poloidal plane and (b,d) the radial structure of the poloidal harmonics, $\umx$, for the peaked and linear $\etai$ profiles respectively. The (orange) shaded region in the poloidal cross-section shows the simulation domain.
 	}
	\label{phifig}
\end{figure}

We test this generalised ballooning theory quantitatively using full 2D numerical solutions to \eqn{model-eqn}.
We first decompose $\phixt$ into poloidal Fourier harmonics, $\phixt=\sum_m \umx\exp\brac{im\theta}$, and then solve the set of coupled equations for the radial dependence of the coefficients, $\umx$, to determine the complex mode frequency, $\omega$, as an eigenvalue of the system.
Radially localised modes are sought and we therefore employ zero Dirichlet boundary conditions in the radial direction.
The results for the two different eigenmode structures are shown in \fig{phifig}.
With the peaked $\etai$ profile the mode is indeed localised on the outboard side, while for the linear profile it is peaked at the top of the plasma.
We find no 2D eigenmode on the outboard side when the $\etai$ profile is linear.
Also shown is the radial dependence of each of the Fourier harmonics.
These mode structures can also be obtained from the 1D ballooning procedure; we do not show them here because they are visibly indistinguishable from these 2D solutions.
The eigenvalues from the 2D analysis are $\omega =-0.025 +i0.316$ and $\omega = -0.110 +i0.239$ for the peaked and linear $\etai$ profiles respectively.
These results are in excellent agreement with the 1D ballooning analysis.

%
%

In \Ref{kim-wakatani} Kim and Wakatani argued that a continuum of modes can exist when $\Omega_x \neq 0$.
However, these are not eigenmodes of the system.
Such a continuum arises because there is a range of solutions to \eqn{env} which provide the desired localisation of $\atz$ and hence the existence of the Fourier transform, \eqn{ft}.
However, in general, those solutions do not provide a form for $A$ that is periodic in $\thz$, which is required for $\phi$ to be periodic in $\theta$.
For this reason, the Kim-Wakatani modes are not physical eigenmodes, as was also noted in \Ref{dewarPRL}.
We do not see Kim-Wakatani modes in our 2D eigenmode solutions, confirming the conclusions of \Ref{dewarPRL}.

\begin{figure}[h]
    \centering
    \resizebox{1.0\columnwidth}{!}{
		\includegraphics[width=0.6\columnwidth]
{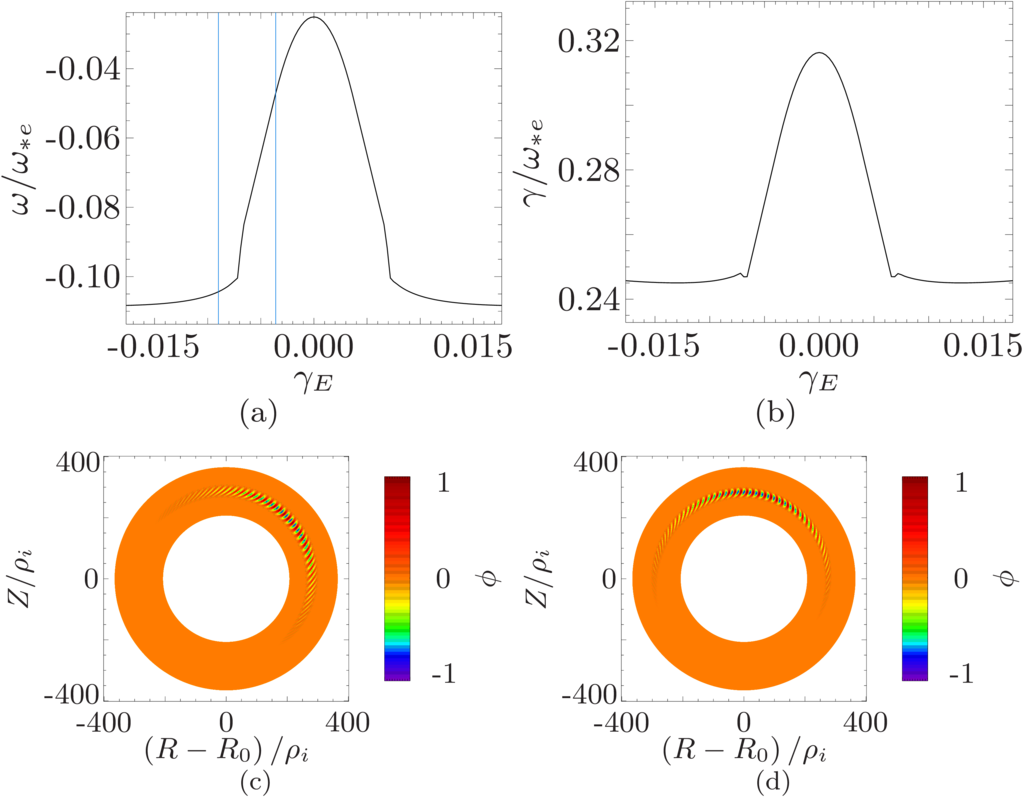}
    	}
 	\caption[]{
	The mode frequency and growth rate (a,b) as a function of the flow shear parameter, $\game$. Also shown are the 2D mode structures for $\game=-0.0036$ (c) and $\game=-0.0090$ (d).
 	}
    \label{flow-scan}
\end{figure}

We can use our 2D solutions to explore the relation between \iso{} and \gen{} modes.
We start with an $\etai$ profile which is peaked at $x=0$ so that the more unstable \iso{} mode exists.
Adding a linear contribution to the $\etai$ profile simply shifts the position of the maximum of $\etai$: the \iso{} mode still exists, but adjusts its position to sit at the point where $\etai$ is a maximum.
A more interesting situation arises when one introduces flow shear into the problem as this then shifts the position where the local frequency is stationary relative to that where the local growth rate peaks.
No isolated mode is then possible.
Thus in \eqn{model-eqn} we introduce a Doppler shift $\omega\rightarrow \omega+\nqp\game x$, working in the rest frame of the rational surface.
The shearing rate, $\game=d\Omega_\varphi/dq$, parameterises the shear in the toroidal rotation, $\Omega_\varphi$.
\Fig{flow-scan}(a,b) shows how the mode frequency and growth rate derived from 2D solutions respond to $\game$; they are symmetric under $\game \rightarrow - \game$.
Note how the \iso{} mode that exists at $\game = 0$, smoothly evolves into the \gen{} mode, as expected from analytic theory \cite{dewarPPCF,connor}, for a relatively low value of $|\game| \sim 0.015\ll\gamma$, where $\gamma$ is the ITG growth rate.
Thus the window of existence for the \iso{} mode is very narrow in $\game$.
We also show in \figs{flow-scan}(c,d) how the eigenmode structure evolves in the poloidal plane from the outboard mid-plane for $\game=0$ (\fig{phifig}(a)) to the top of the plasma at $\game \sim -0.01$.
With a positive value of $\game$ the mode moves towards the bottom of the plasma.
Adding flow shear to a \gen{} mode usually has little impact on stability as the mode type does not change and hence the averaging procedure used in the ballooning analysis to determine the growth rate is unchanged.
The radial width of the mode will, however, be affected because of its dependence on $\Omega_x$.

%
%

In conclusion, we have presented numerical 2D eigenmode solutions for micro-instabilities in toroidal geometry.
These confirm the analytic asymptotic approaches demonstrated in \cite{taywilcon,dewarPPCF}.
In particular we find that the common approach to ballooning theory, selecting the value of $\thz$ to maximise the growth rate which usually gives rise to a mode that balloons on the outboard side, is only valid in very special situations.
More generally, one should average the local (1D ballooning) growth rate over $\thz$, and the dominant mode amplitude will exist off the outboard midplane.
Eigenmode frequencies determined from our 2D calculations for a representative model of toroidal drift waves demonstrate excellent agreement with those determined from the 1D ballooning theory provided one treats $\thz$ correctly.
Considering the effect of flow shear, we have shown that an \iso{} mode smoothly transforms into the more stable \gen{} mode as the flow shear is increased, and we have demonstrated the very narrow region of parameter space where the isolated mode exists.

Our analysis is generic to any micro-instability in a tokamak plasma, and may be relevant to situations where plasma gradients, generally clamped by stiff transport from microturbulence, are transiently relaxed by bursty energy releases.
In the tokamak high confinement-mode, steep density and temperature gradients form a pedestal close to the plasma edge.
The pedestal periodically collapses due to ELMs, with the profiles subsequently rebuilding until the next ELM is triggered.
Large ELMs are believed to be due to the onset of intermediate-$n$ ideal MHD peeling-ballooning modes \cite{peebal, peebal2, eped}, which become unstable for sufficiently high and wide pedestals.
Small ELM regimes are less well understood.
The linear growth rate and mode frequency of the high $n$ micro-instabilities responsible for transport in the steep pedestal region, will not typically be stationary at the same location $\brac{x,\thz}$, so that \iso{} modes will not usually exist.
Turbulence from the more stable \gen{} mode will therefore typically constrain the pedestal gradients.
As profiles rebuild between ELMs, however, the locations where the frequency and growth rate are stationary also evolve and may transiently become collocated, allowing an \iso{} mode to exist.
At that instant the gradient would be close to the \gen{} mode's critical gradient, well above that associated with the \iso{} mode.
Therefore the \iso{} mode would suddenly become highly unstable, potentially triggering a rapid crash in the gradients (i.e. a small ELM).
Such localised crashes of the pedestal profiles may prevent the equilibrium from ever reaching the ideal MHD instability boundary associated with larger type-I ELMs.
This could form the basis of a theory for small ELMs in tokamaks.

\footnotesize
\begin{acknowledgments}
The authors gratefully acknowledge helpful discussions with Bryan Taylor and Jack Connor. 
This work was part-funded by the RCUK Energy Programme under grant EP/I501045 and the European Communities under the contract of Association between EURATOM and CCFE.
To obtain further information on the data and models underlying this paper please contact PublicationsManager@ccfe.ac.uk.
The views and opinions expressed herein do not necessarily reflect those of the European Commission.
\end{acknowledgments}

\bibliographystyle{aip}

\begin{thebibliography}{99}
\bibitem{taywilcon} 
J.B.\ Taylor, H.R.\ Wilson and J.W.\ Connor, \textit{Plasma Phys.\ Control.\ Fusion} {\bf 38} 243 (1996)
\bibitem{dewarPPCF}
R.L.\ Dewar, \textit{Plasma Phys.\ Control.\ Fusion} {\bf 39} 453 (1997)
\bibitem{cht} 
J.W.\ Connor, R.J.\ Hastie and J.B.\ Taylor, \textit{Proc.\ R.\ Soc.\ London Ser.\ A} {\bf 365} 1 (1979)
\bibitem{glasser} 
A.H.\ Glasser in {\it Finite beta theory, Proceedings of the workshop}, Varenna, edited by B.\ Coppi and W.L.\ Sadowski (U.S.\ Dept.\ of Energy, Washington D.C.) CONF-7709167, p.\ 55 (1977)
\bibitem{dewar} 
R.L.\ Dewar and A.H.\ Glasser, \textit{Phys.\ Fluids} {\bf 26} 3038 (1983)
\bibitem{lee} 
Y.C.\ Lee and J.W.\ Van\ Dam in {\it Finite beta theory, Proceedings of the workshop}, Varenna edited by B.\ Coppi and W.L.\ Sadowski (U.S.\ Dept.\ of Energy, Washington D.C.) CONF-7709167, p.\ 93 (1977)
\bibitem{Note1} More generally this position depends on a number of factors including the surface shaping and instability being considered. In a treatment where both the $\Omega_{xx}$ and $\Omega_{x}$ terms are retained the location of peaking will also depend upon the relative size of these two terms allowing the mode to sit between the two limits obtained by neglecting either of these terms, as demonstrated in \fig{flow-scan}.
\bibitem{oyama} N.\ Oyama, P.\ Gohil, L.D.\ Horton, A.E.\ Hubbard, J.W.\ Hughes, Y.\ Kamada, K.\ Kamiya, A.W.\ Leonard, A.\ Loarte, R.\ Maingi, G.\ Saibene, R.\ Sartori, J.K.\ Stober, W.\ Suttrop, H.\ Urano, W.P.\ West and the ITPA Pedestal Topical Group, \textit{Plasma Phys.\ Control.\ Fusion} {\bf 48} A171 (2006)
\bibitem{ct} 
J.W.\ Connor and J.B.\ Taylor, \textit{Phys. Plasmas} {\bf 30} 3180 (1987)
\bibitem{zhang} 
Y.Z.\ Zhang and S.M.\ Mahajan, \textit{Phys.\ Lett.} {\bf 157A} 133 (1991)
\bibitem{kim-wakatani} 
K.Y.\ Kim and M.\ Wakatani, \textit{Phys.\ Rev.\ Lett.} {\bf 73} 2200 (1994)
\bibitem{dewarPRL}
R.L.\ Dewar, Y.Z.\ Zhang and S.M.\ Mahajan, \textit{Phys.\ Rev.\ Lett.} {\bf 74} 4563 (1995)
\bibitem{connor} J.W.\ Connor, R.J.\ Hastie and T.J.\ Martin, ISPP21, Theory of Fusion Plasmas, Varenna, edited by J.W.\  Connor, O.\ Sauter and E.\ Sindoni, 457 (2004)
\bibitem{peebal} H.R.\ Wilson, J.W.\ Connor, A.R.\ Field, S.J.\ Fielding, R.L.\ Miller, L.L.\ Lao, J.R.\ Ferron and A.D.\ Turnbull, \textit{Phys.\ Plasmas} {\bf 6} 1925 (1999)
\bibitem{peebal2} P.B.\ Snyder, H.R.\ Wilson, J.R.\ Ferron, L.L.\ Lao, A.W.\ Leonard, T.H.\ Osborne, A.D.\ Turnbull, D.\ Mossessian, M.\ Murakami and X.Q.\ Xu, \textit{Phys.\ Plasmas} {\bf 9} 2037 (2002)
\bibitem{eped} P.B.\ Snyder, R.J.\ Groebner, A.W.\ Leonard, T.H.\ Osborne and H.R.\ Wilson, \textit{Phys. Plasmas} {\bf 16} 056118 (2009)
\end{thebibliography}

\end{document}